\documentclass[journal=iecred,manuscript=article]{achemso}
\setkeys{acs}{doi=true}

\usepackage{amsmath,bm}
\usepackage[labelformat=simple]{subcaption}
\usepackage{multirow}
\usepackage{siunitx}
\usepackage{orcidlink}

\SectionNumbersOn

\author{Pranay P. Nagrani}
\author{Amy M. Marconnet}
\author{Ivan C. Christov}
\email{christov@purdue.edu}
\affiliation[Purdue University]{School of Mechanical Engineering, Purdue University, West Lafayette, Indiana 47907, USA}

\title[Interphase Drag Correlation for Gas--Liquid Flows]{Correlations for the Interphase Drag in the Two-Fluid Model of Gas--Liquid Flows through Packed-Bed Reactors}

\abbreviations{IR,NMR,UV}
\keywords{American Chemical Society, \LaTeX}

\usepackage{hyperref}
\hypersetup{
     colorlinks=true,
     linkcolor=blue,
     filecolor=blue,
     citecolor=blue,      
     urlcolor=blue,
     breaklinks=true
}

\begin{document}


\begin{abstract}
Experiments conducted by NASA measured the pressure drop due to gas--liquid flow through a packed-bed reactor under microgravity conditions. From these experiments, we develop correlations for the gas--liquid $f_{gl}$ interphase drag in a two-fluid model (TFM). We use an Ergun-type closure for liquid--solid drag. Then, under a 1D flow assumption, $f_{gl}$ is the only unknown in the TFM. Using a data-driven approach, we determine $f_{gl}$ and correlate it (via composite fits) with the liquid and gas Reynolds numbers, $Re_{l}$ and $Re_{g}$, respectively, and the Suratman number $Su_{l}$. To validate the proposed $f_{gl}(Re_{l},Re_{g},Su_{l})$ closure, we perform two-dimensional transient simulations at microgravity conditions using ANSYS Fluent and employing an Euler--Euler formulation. We find good agreement between the simulations based on the proposed $f_{gl}$ closure and the experimental data.
\end{abstract}


\section{Introduction}
\label{sec::Intro}

Multiphase flows through porous media \citep{Adler1988MultiphaseMedia,Wooding1976MultiphaseMedia} find applications across various systems, from heat pipes to volatile removal assembly and hydrogenation plants, to name a few \citep{Azarpour2021PerformanceReview}. One such problem is gas--liquid (two-phase) flow through a packed-bed reactor (PBR). A typical PBR  consists of spherical catalyst particles packed within a cylindrical (or cuboidal) pipe \citep{Ranade2011TrickleApplications}. In the presence of gravity, the gas and liquid phases arrange themselves in the trickle flow regime (the liquid phase trickles over the packing in a gas continuous medium), and the corresponding pressure drops and trickle-to-pulse flow regime transitions have been studied extensively (see, for instance, Refs.~\citenum{Wang2013ModellingReview,Attou1999ModellingReactor,Holub1992AFLOW,Holub1993PressureFlow} and those therein). On the other hand, in the absence of gravity (or under microgravity conditions), the capillary force becomes important. Bubble flow (gas droplets in liquid continuous medium) and pulse flow (alternating regions of gas-rich and liquid-rich) regimes are observed in two-phase flow through PBRs at low-gravity conditions \citep{Motil2021GasliquidExperiment,Motil2003GasliquidMicrogravity,Taghavi2022GasliquidExperiment2}. However, determining the pressure drop across the PBR at microgravity conditions remains an active research problem \citep{Balasubramaniam2006TwoGravity}. Recently, progress has been made via experimental and modeling efforts.

Specifically, \citet{Motil2003GasliquidMicrogravity} reported on parabolic flight experiments of gas--liquid flow through a PBR. By developing flow regime maps, they demonstrated that different flow physics compete in microgravity conditions compared to terrestrial gravity conditions. In addition, a correlation known as the ``modified Ergun equation'' was proposed to predict the frictional pressure drop for gas--liquid flows through PBRs at microgravity conditions. The duration of low-gravity conditions on parabolic flights is short. To remedy this shortcoming and gather data on longer-duration PBR flows, further experiments were conducted aboard the International Space Station (ISS)\citep{Motil2021GasliquidExperiment,Taghavi2022GasliquidExperiment2}. Specifically, the wettability and hysteresis effects were studied in the viscous-capillary regime (low Reynolds number flow) by measuring pressure drops and bubble-to-pulse flow regime transitions for Teflon (nonwetting) and glass (wetting) packing materials \citep{Motil2021GasliquidExperiment}. These experiments suggested that contributions from capillarity dominate the pressure drop for wetting packing, whereas contributions from viscous drag dominate the pressure drop for the nonwetting cases. Further, the effect of preflow (i.e., gas-flush versus liquid-flush) through a PBR on the pressure drop was investigated  \citep{Taghavi2022GasliquidExperiment2}. Higher gas holdup and pressure drops were observed for liquid preflow conditions compared to gas preflow conditions. More recently, a technique to characterize flow regime transitions via the slope of the pressure drop versus flow rate curve (from the ISS data) demonstrated the dominance of capillarity at low flow rates and the dependence of the gas holdup on preflow conditions \citep{Taghavi2023TheFlows}. 

One-dimensional (1D) two-fluid models (TFMs) have been developed to understand these experimental observations from a fundamental (continuum) point of view \citep{Gidaspow1994MultiphaseDescriptions}. Assuming the flow to be 1D, uniform in the cross-section, and at steady-state, the full TFM equations can be simplified so that the unknowns are the axial pressure drop, the liquid holdup, and the interphase drag force \citep{Zhang2017HydrodynamicsModel,Salgi2014ImpactBeds,Salgi2017Experimentally-basedBeds}. Following this approach,  \citet{Zhang2017HydrodynamicsModel} performed experiments and developed a 1D TFM at low Bond number $Bo$ (the ratio of gravitational to surface tension forces), within which the trickle flow regime is dominant. They incorporated Ergun-type correlations in the three interphase drag forces (i.e., the liquid--solid $f_{ls}$, gas--liquid $f_{gl}$, and gas--solid $f_{gs}$ ones) with appropriate relative velocities \citep{Attou1999ModellingReactor,Holub1992AFLOW,Holub1993PressureFlow}. At low $Bo$, capillarity implies a pressure jump (drop) across a gas--liquid interface: $p_c=p_g-p_l\ne0$, where $p$ is pressure and the subscripts ``$c$", ``$g$", and ``$l$" stand for ``capillary,'' ``gas'', and ``liquid,'' respectively. For example, \citet{Zhang2017HydrodynamicsModel} employed the correlation from \citet{Attou2000AReactor} to model $p_c$ as a function of surface tension, particle diameter $d_p$, gas holdup, and the ratio of gas and liquid densities. Next, estimating $d p_c/d z$ as $p_c/d_p$, where $p_c$ comes from the previous correlation, the respective pressures in the 1D gas and liquid phases' momentum equations were eliminated to obtain a single momentum equation, with the liquid holdup being the only unknown. Finally, the pressure drop across the PBR was calculated by substituting the predicted liquid holdups, which were determined for different gas and liquid flow rates, obtaining a good agreement between the 1D TFM predictions and the experiments. 

Meanwhile, \citet{Salgi2014ImpactBeds} used a modified Ergun correlation to calculate the pressure drop under microgravity conditions \citep{Motil2003GasliquidMicrogravity}. They treated the liquid holdup and $f_{gl}$ as unknown in their model. They used a so-called ``dynamic interaction term'' from \citet{Motil2003GasliquidMicrogravity} to improve the fit to the experimental pressure drops. Further, a capillary pressure drop correlation, derived under terrestrial gravity conditions, was incorporated because there are no acceptable capillary pressure drop correlations for microgravity conditions. A linear stability analysis was performed to obtain $f_{gl}$ and liquid holdups at different gas and liquid flow rates. Similarly, \citet{Salgi2017Experimentally-basedBeds} developed experimentally-based correlations for $f_{gl}$ from pulse velocity and frequency data to predict pressure drops through PBRs at both microgravity and terrestrial gravity conditions. 

In these prior 1D TFM studies, model calibration (of one or more fitting parameters) is key to accurately predicting pressure drops. Further, as \citet{Motil2021GasliquidExperiment} stated in their paper, ``Ergun type expressions for $f_{gl}$ cannot be justified in the dispersed bubble flow regime." However, it is the dispersed bubble and pulse flow regimes that are observed at microgravity conditions \citep{Motil2021GasliquidExperiment,Motil2003GasliquidMicrogravity}, which we seek to model in this work. Therefore, the goal of the present work is to perform a data-driven calculation of the drag coefficient (fitting coefficient of our model) to obtain composite fits that yield a correlation for $f_{gl}$ at microgravity conditions. 

In addition to theoretical efforts, computational fluid dynamics (CFD) has been widely used to study resolved two-phase flows through PBRs. Specifically, an Euler--Euler TFM approach is used to predict pressure drop and liquid holdup within PBRs. Previous CFD studies\citep{Atta2007PredictionCFD,Atta2007InvestigationCFD,Lopes2008Three-dimensionalReactor,Lu2018ABeds,Gunjal2005HydrodynamicsModeling} considered an ``empty'' channel and incorporated the relevant closure relations for interphase drag terms (such as $f_{gl}$, $f_{ls}$, and $f_{gs}$) as momentum source terms, which implicitly accounts for the porous medium. In this modeling approach, one does not have to resolve the complex flow domain of interstitial void spaces to extract relevant flow parameters \citep{Nagrani2024HydrodynamicsReactors}. In the present study, we follow this approach and incorporate the proposed $f_{gl}$ closure relations (developed from the 1D TFM model) as momentum source terms in CFD simulation. Then, we compare the pressure drop from these CFD simulations with the experimental pressure drop measured by NASA's Packed Bed Reactor Experiment (PBRE), which accessed only the bubble and pulse flow regimes at microgravity, provided as data in NASA's Physical Sciences Informatics (PSI) system \citep{Motil2020PackedPBRE2}. The PBRE data in the PSI repository \citep{Motil2020PackedPBRE2} consists of two datasets: one for a glass packing (wetting material) and one for a Teflon packing (nonwetting material). For each dataset, the gas and liquid flow rates were fixed (controlled), and the pressure gradient across the PBR was measured. The comparison between the CFD simulations and PBRE data serves as an example demonstration of the utility of the proposed $f_{gl}$ correlations.

This paper is organized as follows. In Sec.~\ref{sec::math_model}, we introduce the governing equations (Sec.~\ref{sec:cons_law}) and modeling approach (Sec.~\ref{sec:interphase_force}). In Sec.~\ref{sec:calibration}, we discuss the data-driven development of correlations for $f_{gl}$. Then, in Sec.~\ref{sec::cfd_model}, we showcase the utility of the proposed $f_{gl}$ correlation in a CFD simulation (Sec.~\ref{sec:comp_method}) for two different scenarios --- wetting and nonwetting packing materials, and we compare the pressure drop calculated by CFD against the experimental ones (Sec.~\ref{sec:pressure_drop}). Section~\ref{sec::Conclusion} concludes the study.

\section{Mathematical Model}
\label{sec::math_model}
Our specific aim is to obtain a closure relation for interphase gas--liquid drag force $f_{gl}$ in a TFM for flow through a PBR, based on the PBRE microgravity experiments \citep{Motil2020PackedPBRE2}. The idea of a TFM is to treat the carrier and dispersed phases of a two-phase flow as interpenetrating continua, each having its own thermophysical properties and obeying suitable balance laws \citep{Gidaspow1994MultiphaseDescriptions}. The two continua ``interact'' through the interphase drag force, denoted $f$. Therefore, the interphase drag force couples the momentum equations of the two phases. Throughout, it should be understood that $f$ usually represents a body force density (i.e., a force per unit volume), but we refer to it as ``force'' for clarity and simplicity.

In this section, we present our proposed procedure for obtaining $f_{gl}$. Specifically, in Sec.~\ref{sec:cons_law}, we start from the basic mass and momentum balances under the TFM. In Sec.~\ref{sec:interphase_force}, we employ experimental data from NASA's PBRE \citep{Motil2020PackedPBRE2} to perform a data-driven inference and calibration of a model for $f_{gl}$. 

\subsection{Governing Equations}
\label{sec:cons_law}
The governing equations of gas--liquid flow through a PBR are those of conservation of mass and conservation of momentum. Following the predominant approach in the existing literature \citep{Salgi2014ImpactBeds,Salgi2017Experimentally-basedBeds,Zhang2017HydrodynamicsModel}, and in the absence of detailed measurements of the interstitial flow and interface dynamics, the flow is considered predominantly 1D, with only the gradients along the axial (i.e., flow) direction of PBR being of interest in predicting the pressure drop. Under this assumption, the conservation of mass for liquid and gas phases, respectively, can then be simplified to:
\begin{subequations}\begin{align}
    \frac{\partial \phi_l}{\partial t} + v_l  \frac{\partial \phi_l}{\partial z} + \phi_l  \frac{\partial v_l}{\partial z} &= 0,\\
     -\left(\frac{\partial \phi_l}{\partial t} + v_g  \frac{\partial \phi_l}{\partial z}\right) + \phi_g  \frac{\partial v_g}{\partial z} &= 0,
\end{align}\label{eq::com_1d}\end{subequations}
where $t$ is time, $z$ is the axial coordinate, $\phi(z,t)$ is the volume fraction (and $\phi_l$ is termed the  \emph{liquid holdup}), and $v(z,t)$ is the mesoscale velocity. Here, the subscripts ``$g$", ``$l$", and later ``$s$", stand for ``gas,'' ``liquid,'' and ``solid'' phases, respectively.

In a 1D TFM model of two-phase flow through a PBR, it is generally assumed (an approximation is made) that the mesoscale velocities are   $v_g = v_{gs}/(\epsilon\phi_g)$ and $v_l = v_{ls}/(\epsilon\phi_l)$, where $v_{gs}$ and $v_{ls}$ are the corresponding gas or liquid superficial velocities, and $\epsilon$ is the PBR void fraction (porosity) \citep{Salgi2014ImpactBeds,Salgi2017Experimentally-basedBeds,Zhang2017HydrodynamicsModel}. Next, the 1D conservation of momentum equations for the liquid and gas phases (based on the mesoscale velocities), respectively, are
\begin{subequations}\begin{align}
    \rho_l \frac{Dv_l}{Dt} &= -\phi_l\frac{\partial p_l}{\partial z}+\mu_{l}\frac{\partial}{\partial z}\left(\phi_l\frac{\partial v_l}{\partial z}\right) + f_{l} + \rho_l g,\\
    \phi_g \left(\rho_g \frac{Dv_g}{Dt}-\rho_l \frac{Dv_l}{Dt}\right) &= -\phi_g\frac{\partial p_g}{\partial z} + f_{g} + \phi_g(\rho_g-\rho_l) g,
\end{align}\label{eq::colm_1d}\end{subequations}
where $D/Dt$ is the material derivative, $\rho$ is density, $p$ is pressure, $\mu$ is dynamic viscosity, $g$ is the acceleration due to gravity, and $f$ represents body forces densities. The gas phase is treated as inviscid. 

To close the TFM (Eqs.~\eqref{eq::com_1d} and \eqref{eq::colm_1d}), we need to specify the interphase momentum transfer terms, which are represented by the forces $f_g$ and $f_l$. Note that the expressions for $f_g$ and $f_l$ can each have a number of additive contributions, depending on the physics at hand (such as virtual mass, lift, etc.). In the present work, we focus on the main contribution, i.e., the drag. To arrive at expressions for $f_g$ and $f_l$, we assume that the interaction between the gas phase and the solid packing is negligible, i.e., $f_{gs}=0$, because the liquid wets the solid packing \citep{Salgi2014ImpactBeds,Salgi2017Experimentally-basedBeds}. Then, we write the phase-specific forces in terms of the interphase ones as $f_l = f_{gl}-f_{ls}$ and $f_g = -f_{gl}$. We remind the reader that $f_{ls}$ is the interphase drag force between the liquid and solid phases, and $f_{gl}$ is the interphase drag force between the gas and liquid phases. 

Next, it is standard to assume fully developed flow at steady state \citep{Salgi2014ImpactBeds,Salgi2017Experimentally-basedBeds,Zhang2017HydrodynamicsModel} and that the pressure gradients within the gas and liquid phases are equal \cite{Holub1992AFLOW, Holub1993PressureFlow, Attou1999ModellingReactor}. Incorporating these assumptions into Eqs.~\eqref{eq::colm_1d}, we have
\begin{subequations}\begin{align}
    -\phi_l \left.\frac{dp}{dz}\right|_\text{total} - f_{ls} + f_{gl} + \rho_l g = 0,\label{eqn::colm_l}\\
    -\phi_g \left.\frac{dp}{dz}\right|_\text{total} - f_{gl} + (\rho_g-\rho_l)\phi_g g = 0. \label{eqn::colm_g}
\end{align}\label{eqn::colm_assumption}\end{subequations}

Note that the static pressure contribution (due to gravity) is negligible in microgravity conditions. In such cases, the total pressure gradient is essentially only due to the frictional contribution (from viscous stresses). Therefore, for ease of representation, we omit the subscript ``total" on $-dp/dz$. 

\subsection{Interphase Drag Forces}
\label{sec:interphase_force}
Next, we construct correlations for the interphase drag forces, $f_{ls}$ in Sec.~\ref{sec:fls} and $f_{gl}$ in Sec.~\ref{sec:fgl}, required to close Eqs.~\eqref{eqn::colm_assumption}. 

\subsubsection{Two-Phase Ergun Correlation for the Liquid--Solid Drag}
\label{sec:fls}
Previous studies\citep{Salgi2014ImpactBeds,Salgi2017Experimentally-basedBeds} used an Ergun-type  \citep{Ergun1949FluidBeds,Ergun1952FluidColumns} equation to determine $f_{ls}$. An Ergun-type equation consists of additive viscous and inertial contributions to the pressure gradient across a PBR, but the relation can also be used to express the drag. Specifically,  \citet{Salgi2014ImpactBeds,Salgi2017Experimentally-basedBeds} adopted the Ergun-type equation:
\begin{equation}
    f_{ls} = \frac{1}{\phi_l^2} \underbrace{\left[\frac{180(1-\epsilon)^2\mu_l v_{ls}}{\epsilon^3 d_p^2} + \frac{1.8(1-\epsilon)\rho_l v_{ls}^2}{\epsilon^3 d_p}\right]}_{=A_{ls}} ,
    \label{eq::f_ls}
\end{equation}
where $d_p$ is the diameter of the spherical particles comprising the packing. Here, $f_{ls}=A_{ls}$ is a single-phase Ergun-type  \citep{Ergun1949FluidBeds,Ergun1952FluidColumns} equation for the liquid phase, which has been adjusted by the denominator $\phi_l^2$ to account for the presence of a gas phase reducing the area available for the flow of the liquid phase. In the present study, we use Eq.~\eqref{eq::f_ls} to model the drag between the liquid and solid phases where the coefficients $180$ and $1.8$ for vicious and inertia components respectively are from \citet{Motil2021GasliquidExperiment}.

\subsubsection{Formulation for the Gas--Liquid Drag}
\label{sec:fgl}
Next, we describe our approach to obtaining the closure relation (i.e., correlation) for $f_{gl}$ from the experimental pressure drop data obtained from NASA's PBRE, which we accessed via NASA's PSI \citep{Motil2020PackedPBRE2}. As indicated in the works of \citet{Salgi2014ImpactBeds}, bubble and pulse flow regimes are predominant at microgravity conditions, and the trickle flow regime does not occur. Hence, one cannot use an Ergun-type correlation for $f_{gl}$, with a relative velocity because that is valid only in the trickle flow regime \citep{Attou1999ModellingReactor,Holub1992AFLOW,Holub1993PressureFlow}. To circumvent this issue, we use a traditional drag force formulation: $f_{gl}=\beta (v_g-v_l)$, where  $\beta$ is the drag coefficient, and $v_g-v_l$ is the relative velocity. Using this formulation, the shared pressure gradient can be eliminated between Eqs.~\eqref{eqn::colm_l} and \eqref{eqn::colm_g}, while $f_{ls}$ is eliminated via Eq.~\eqref{eq::f_ls}, to obtain a single governing equation:
\begin{equation}
    (\rho_l-\rho_g)g + \frac{\beta (v_g - v_l) - \frac{A_{ls}}{\phi_l^2} + \frac{A_{ls}}{\phi_l}}{\phi_g \phi_l} + \frac{\rho_l g}{\phi_l} = 0.
    \label{eq::colm_combined}
\end{equation}

\citet{Salgi2014ImpactBeds} noted the absence of an acceptable correlation for the capillary pressure $p_c$ at microgravity conditions. While previous work \cite{Zhang2017HydrodynamicsModel} estimated $dp_c/dz$ as $p_c/d_p$ using a correlation based on terrestrial gravity conditions for $p_c$, we follow the reasoning of \citet{Attou2000AReactor}, namely ``for a fully established gas--liquid flow, the gradient of the capillary pressure can be reasonably neglected compared with the gradient of pressure in each fluid,'' justifying the shared pressure gradient assumption that led to Eq.~\eqref{eq::colm_combined}.

\section{Data-Driven Calibration of the Mathematical Model}
\label{sec:calibration}

\subsection{Description of the PBRE Dataset}

The PBRE data in the PSI repository \citep{Motil2020PackedPBRE2} consists of two sets, one for glass packing (wetting) and one for Teflon packing (nonwetting). Each dataset has gas and liquid flow rates as input and the resulting pressure gradient across the PBR as output. The absolute pressure was also measured at five axial locations along the PBR. In the present analysis, we exclude the data in the PBRE data that was explicitly labeled as having significant experimental errors, such as problems with the setup and measurements done aboard the ISS for these cases. We repeat our analysis shown below twice for each of the packing materials used in the PBRE to implicitly capture the contact angle dependence (effect of wetting) in our model. 

To use the PBRE data in conjunction with  Eq.~\eqref{eq::colm_combined}, the flow rates must be converted to $v_{gs}$ and $v_{ls}$ and $-dp/dz$ into $\phi_l$. First, the flow rates can be converted into relevant gas and liquid mass fluxes (units of $\si{\kilo\gram \per\meter\square \per\second}$) represented as $\dot{m}''_g$ and $\dot{m}''_l$, then  $v_{ls}=\dot{m}''_l/\rho_l$ and  $v_{gs}=\dot{m}''_g/\rho_g$ are the corresponding superficial velocities. We further define dimensionless Reynolds numbers based on the superficial velocities as: $Re^*_{gs}=\rho_g v_{gs} d_p / \mu_g (1-\epsilon)$ and $Re^*_{ls}=\rho_l v_{ls} d_p / \mu_l (1-\epsilon)$. Next, as explained by \citet{Salgi2014ImpactBeds}, $-dp/dz$ and $\phi_l$ can be related as $\phi_l=\sqrt{A_{ls}/(-dp/dz)}$ by adding Eqs.~\eqref{eqn::colm_assumption} together and eliminating the $f_{ls}$ via Eq.~\eqref{eq::f_ls}. Hence, the $-dp/dz$ data from PBRE datasets can be converted into $\phi_l$.

\subsection{Calibration Procedure}

\begin{figure}[t]
    \begin{subfigure}[t]{\linewidth}
      \centering
      \includegraphics[width=0.85\linewidth]{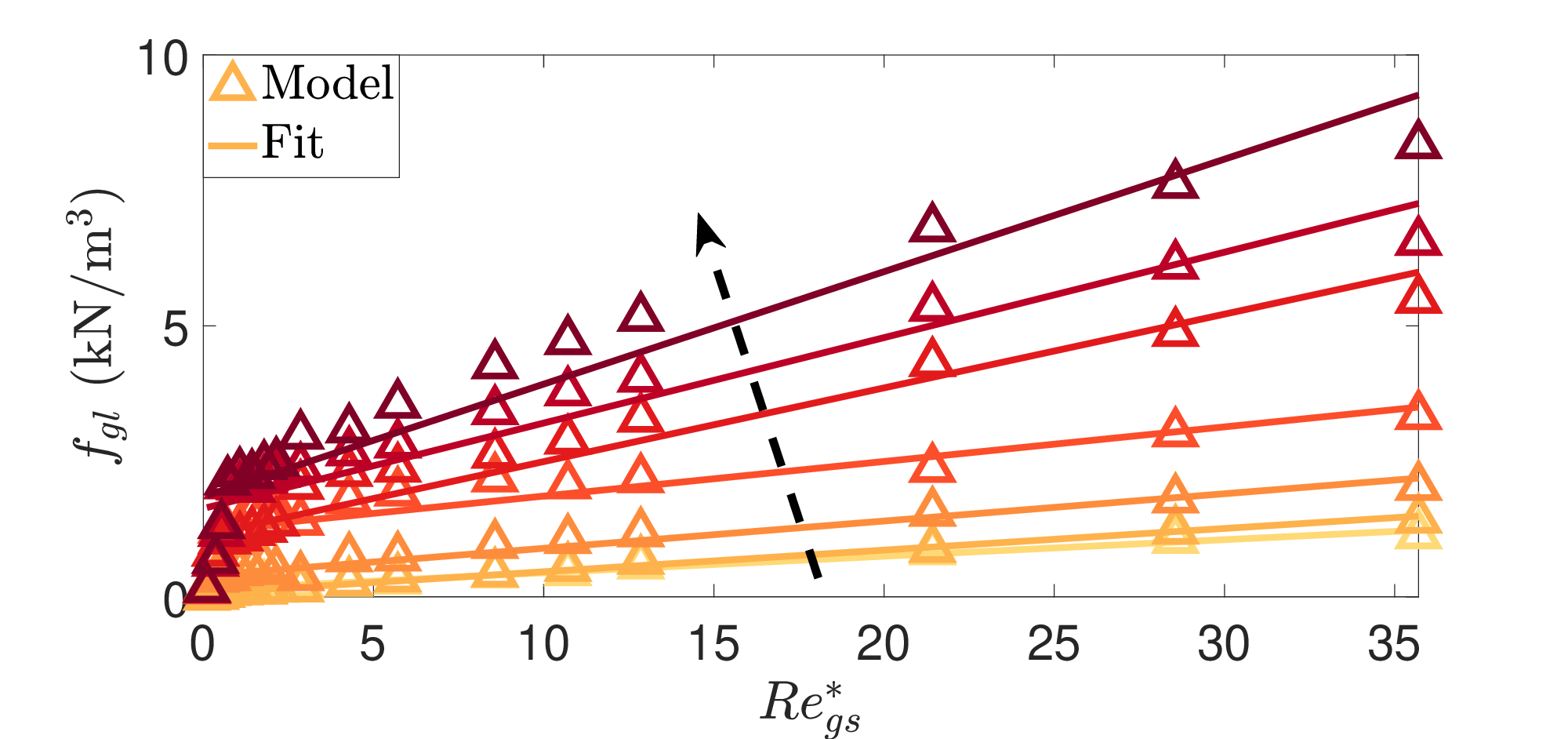} 
      \caption{}
      \label{fig:fgl_fit_teflon}
    \end{subfigure}
    \begin{subfigure}[]{0.45\linewidth}
      \centering
      \includegraphics[width=\linewidth]{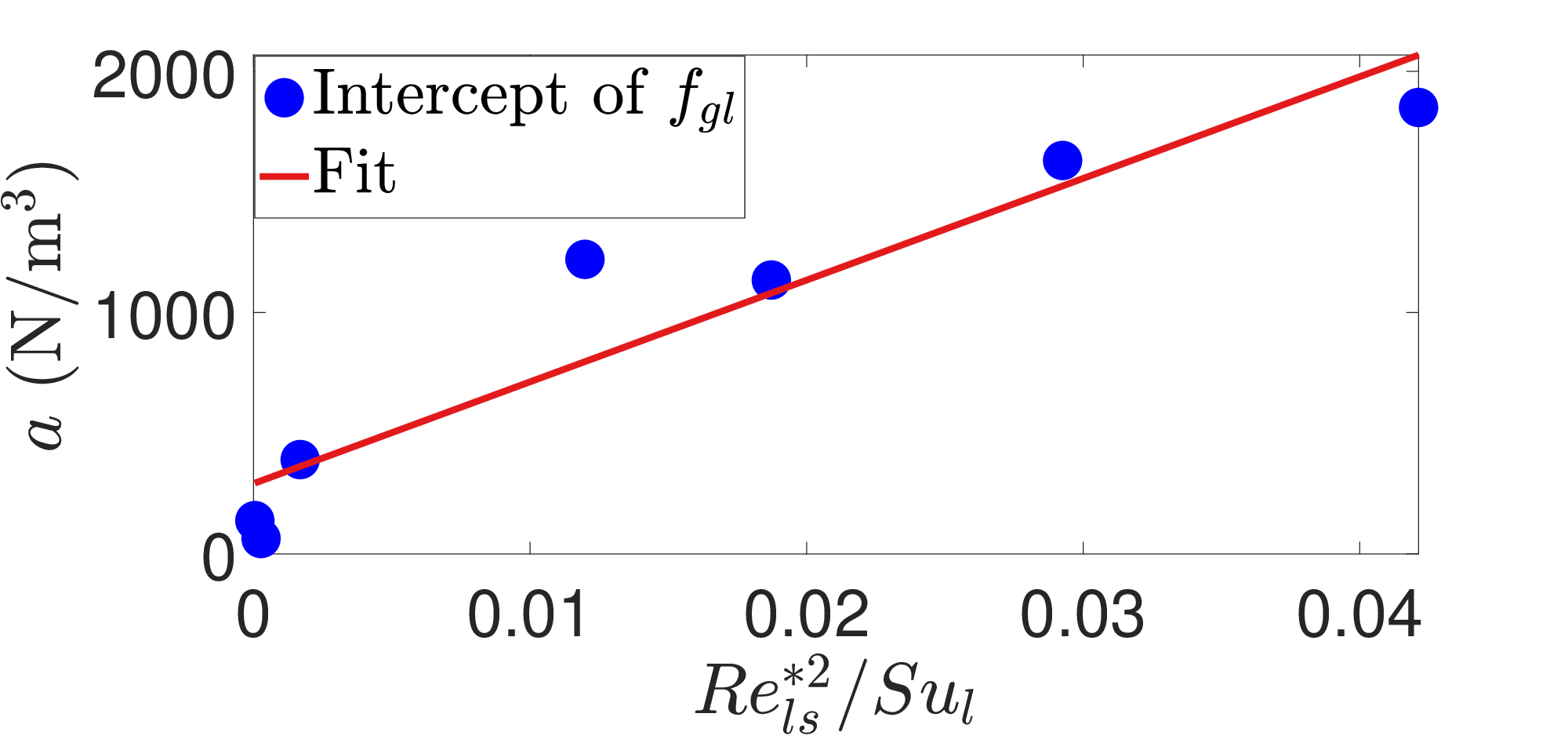}
      \caption{}
      \label{fig:a_fit_teflon}
    \end{subfigure}
    \begin{subfigure}[]{0.45\linewidth}
      \includegraphics[width=\linewidth]{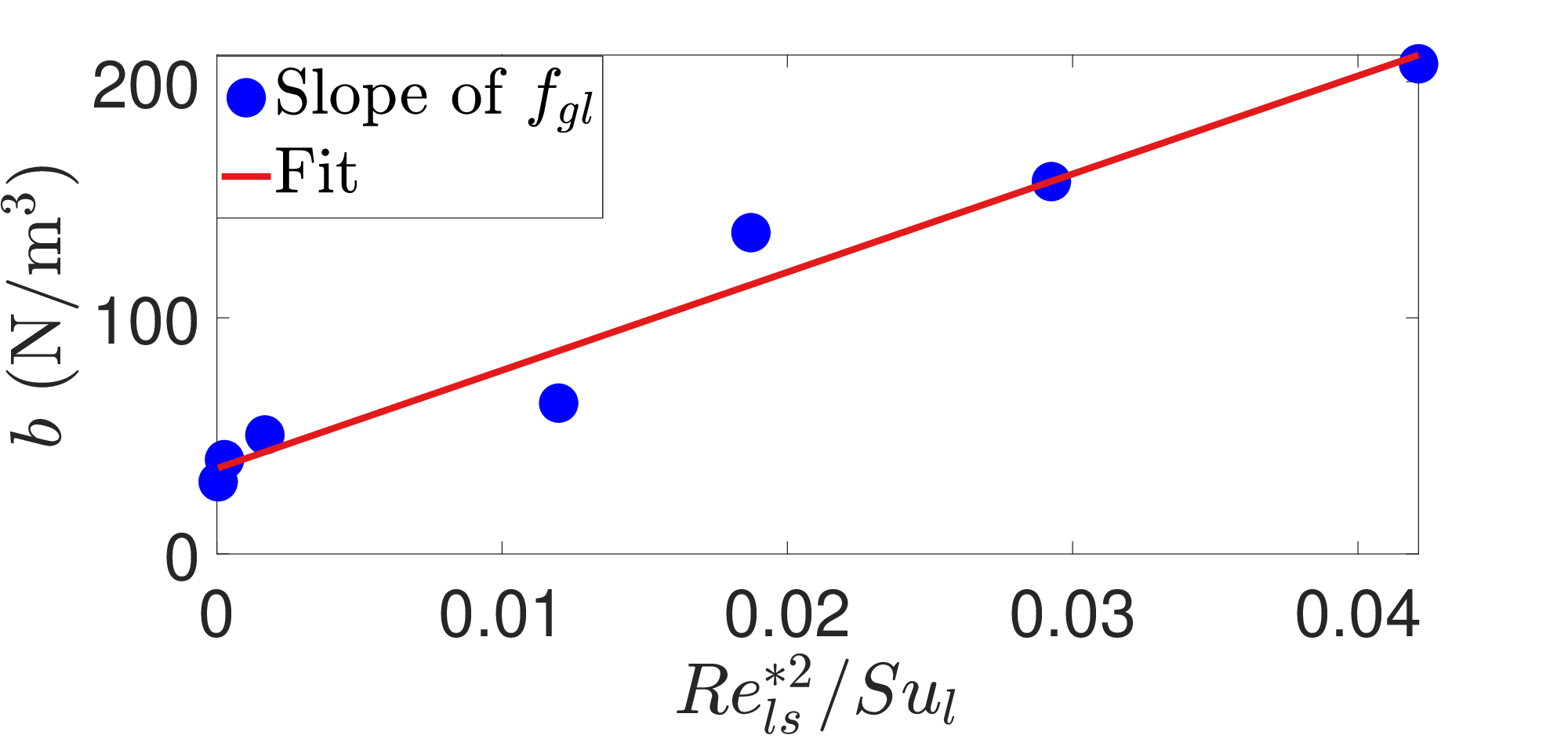}
      \caption{}
      \label{fig:b_fit_teflon}
    \end{subfigure}
    \caption{(a) Variation of $f_{gl}$ with $Re^*_{gs}$ for the Teflon packing and corresponding linear fits of the form: $f_{gl}(Re^*_{ls},Re^*_{gs})=a(Re^{*2}_{ls}/Su_{l})+b(Re^{*2}_{ls}/Su_{l})Re^*_{gs}$. The values of $Re^*_{ls}$ (represent by different colors) used are $3.5$, $8.4$, $20.9$, $55.7$, $69.6$, $87.0$, and $104.4$. The arrow represents the direction of increasing $Re^*_{ls}$. (b) Linear fit for the intercepts of $f_{gl}$ of the form: $a(Re^{*2}_{ls}/Su_{l})=\alpha+\beta Re^{*2}_{ls}/Su_{l}$. (c) Linear fit for the slopes of $f_{gl}$ of the form: $b(Re^{*2}_{ls}/Su_{l})=\gamma+\delta Re^{*2}_{ls}/Su_{l}$.}
    \label{fig::fgl_fit_teflon_all}
\end{figure}

Using $v_{ls}$ and $v_{gs}$ as inputs and $\phi_l$ as output, and along with the PBR's properties, Eq.~\eqref{eq::colm_combined} has only $\beta$ as an unknown. We used the \texttt{lsqnonlin} function of  \textsc{Matlab} to calculate $\beta$ from Eq.~\eqref{eq::colm_combined} for \emph{each} set of conditions processed from the PBRE datasets. Specifically, we calculated $\beta$ for each set of input velocities. Then, turning the input velocities into Reynolds numbers, we calculated $f_{gl}=\beta (v_g - v_l)$ as a function of the input $Re^*_{gs}$ and $Re^*_{ls}$ for the Teflon packing dataset. Finally, we plot $f_{gl}$ versus $Re^*_{gs}$ in Fig.~\ref{fig::fgl_fit_teflon_all} and observe a linear dependence for each $Re^*_{ls}$ (represented by color).

\begin{figure*}[ht]
    \begin{subfigure}[t]{\linewidth}
      \centering
      \includegraphics[width=0.85\linewidth]{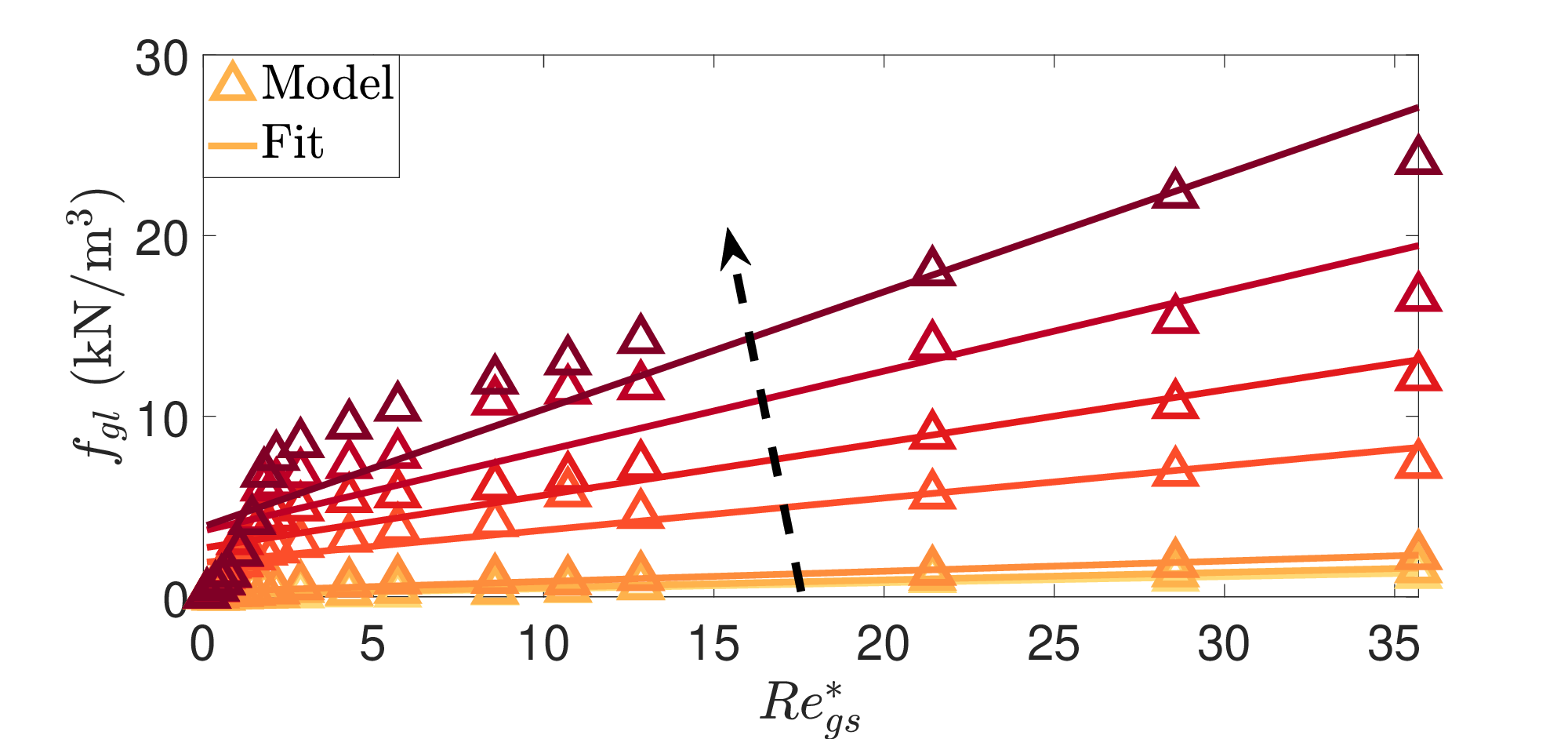} 
      \caption{}
      \label{fig:fgl_fit_glass}
    \end{subfigure}
    \begin{subfigure}[t]{0.45\linewidth}
      \centering
      \includegraphics[width=\linewidth]{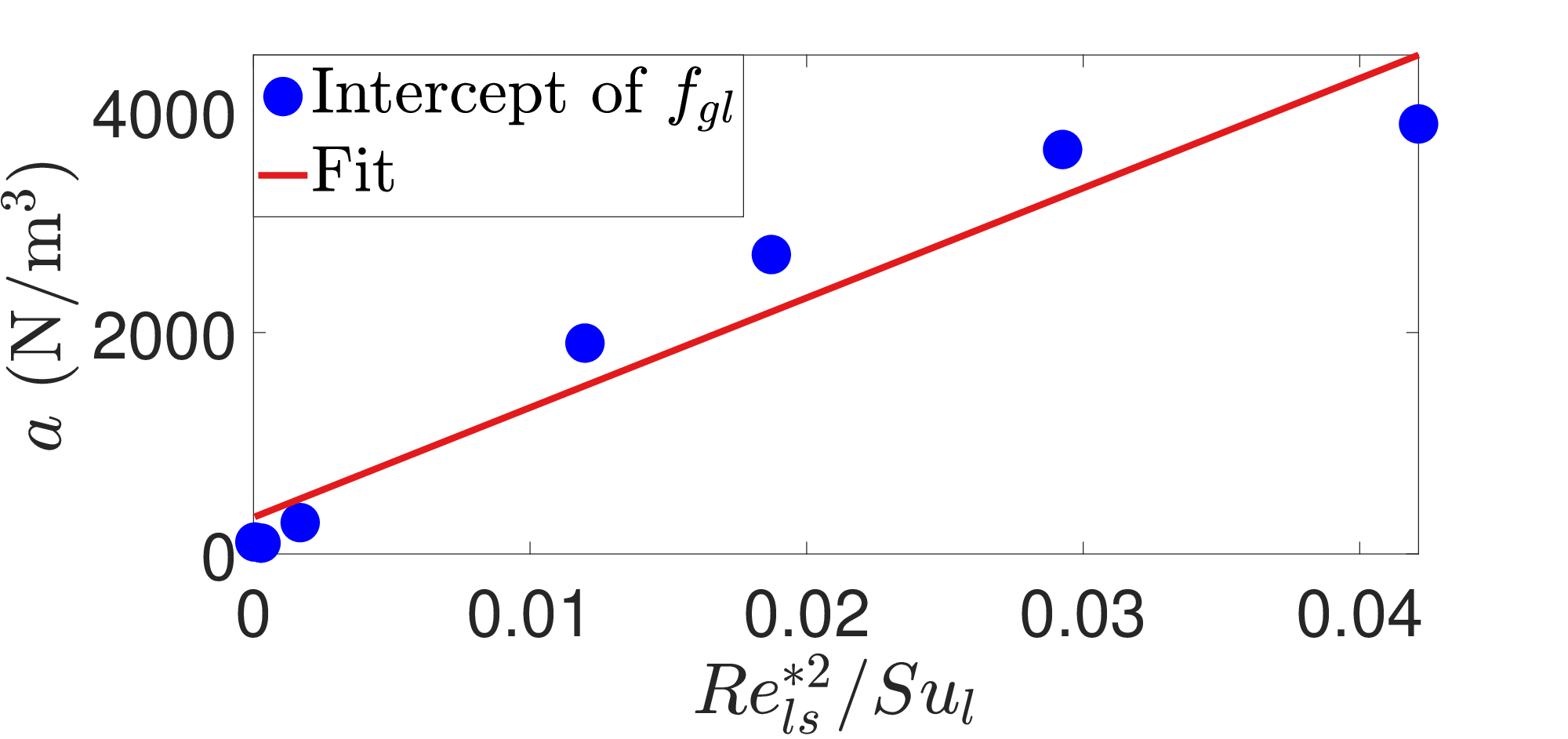}
      \caption{}
      \label{fig:a_fit_glass}
    \end{subfigure}
    \begin{subfigure}[t]{0.45\linewidth}
      \includegraphics[width=\linewidth]{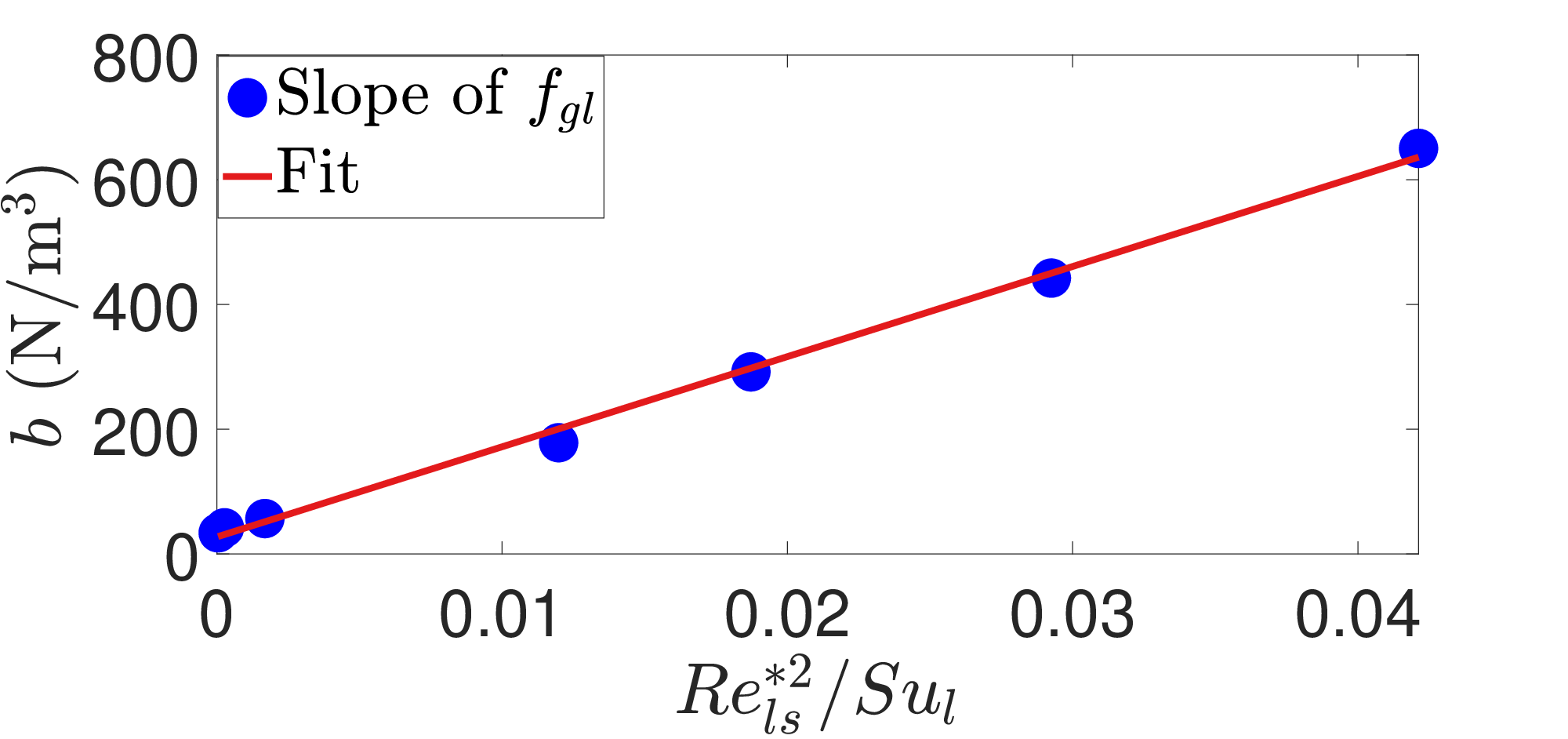}
      \caption{}
      \label{fig:b_fit_glass}
    \end{subfigure}
    \caption{(a) Variation of $f_{gl}$ with $Re^*_{gs}$ for the glass packing and corresponding linear fits of the form: $f_{gl}(Re^*_{ls},Re^*_{gs})=a(Re^{*2}_{ls}/Su_{l})+b(Re^{*2}_{ls}/Su_{l})Re^*_{gs}$. The values of $Re^*_{ls}$ (represent by different colors) used are $3.5$, $8.4$, $20.9$, $55.7$, $69.6$, $87.0$, and $104.4$. The arrow represents the direction of increasing $Re^*_{ls}$. (b) Linear fit for the intercepts of $f_{gl}$ of the form: $a(Re^{*2}_{ls}/Su_{l})=\alpha+\beta Re^{*2}_{ls}/Su_{l}$. (c)  Linear fit for the slopes of $f_{gl}$ of the form: $b(Re^{*2}_{ls}/Su_{l})=\gamma+\delta Re^{*2}_{ls}/Su_{l}$.}
    \label{fig::fgl_fit_glass_all}
\end{figure*}

To quantify this dependence, we performed a set of linear curve fits on the $f_{gl}$ versus $Re^*_{gs}$ data at each $Re^*_{ls}$ to obtain correlations of the form: $f_{gl}=a+b Re^*_{gs}$, where $a$ and $b$ are, respectively, the intercept and slope of the curvefit at fixed value of $Re^*_{ls}$. To combine these multiple correlations into a single correlation, we performed a composite linear fit to the coefficients $a$ and $b$ (obtained across all $Re^*_{ls}$ values) with respect to $Re^{*2}_{ls}/Su_{l}$, where $Su_l = \rho_l d_p \sigma / \mu_l^2$ is the Suratman number, as shown in Fig.~\ref{fig:a_fit_teflon} and Fig.~\ref{fig:b_fit_teflon} of the forms: $a(\xi)=\alpha+\beta \xi$ and $b(\xi)=\gamma+\delta\xi$. We introduced $Su_{l}$ (a combined measure of the relative importance of surface tension to both inertial and viscous forces in the flow) here to implicitly account for surface tension (capillarity effects) between the gas and liquid phases, even if the capillary pressure gradient can be neglected in fully developed steady flow \cite{Attou2000AReactor}. Further, following the ideas leading \citet{Motil2003GasliquidMicrogravity} to their dynamic interaction term, we correlate $a$ and $b$ as functions of $Re^{*2}_{ls}/Su_{l} \sim We_{l}$ (the Weber number). Figure~\ref{fig:fgl_fit_teflon} shows that for largest values of $Re^*_{ls}$, the dependence of $f_{gl}$ on $Re^*_{gs}$ tends to be nonlinear for small values of $Re^*_{gs}$. However, since we performed a curve fit over the full range of data, these outliers do not significantly affect the trend and are not handled differently (see also the discussion of Fig.~\ref{fig::model_val} below). In principle, fitting functions with more parameters could be devised to capture those points better, but given the uncertainties involved at low flow rates, such an approach would likely result in \emph{overfitting}.

We performed an identical analysis on the PBRE dataset for a glass packing. The corresponding data and curve fit are shown in Fig.~\ref{fig::fgl_fit_glass_all}.   

Finally, the proposed interphase drag force closure relation is $f_{gl}=a(Re^{*2}_{ls}/Su_{l}) + b(Re^{*2}_{ls}/Su_{l})Re^*_{gs}$. Explicitly, the correlations obtained for the Teflon and glass packings, respectively, are
\begin{subequations}\begin{align}
    f_{gl}^\text{Teflon} &= 291.6 + 4.22\times10^4 \frac{Re^{*2}_{ls}}{Su_{l}} + 36.2 Re^*_{gs} + 4517 \frac{Re^{*2}_{ls} Re^*_{gs}}{Su_{l}},\\
    f_{gl}^\text{glass} &= 330.8 + 9.92\times10^4 \frac{Re^{*2}_{ls}}{Su_{l}} + 27.2 Re^*_{gs} + 1.466\times10^4 \frac{Re^{*2}_{ls} Re^*_{gs}}{Su_{l}}.
\end{align}\label{eqn::f_gl}\end{subequations}

Finally, if one were to justify that the capillary pressure gradient in a two-phase flow in microgravity conditions is not negligible, it can be added back into Eq.~\eqref{eq::colm_combined}, potentially via a correlation of the form $p_c/d_p$ \cite{Zhang2017HydrodynamicsModel}.  In such a case, the fitting process can be repeated, and the curve-fit coefficients would change, but the methodology remains the same. It must also be noted that the proposed correlation for $f_{gl}$ is applicable for monodisperse Teflon and glass packing materials and cannot be generalized for other packing materials with different sizes. This drawback stems from the limited nature of the experimental data from NASA's PBRE \citep{Motil2020PackedPBRE2}. Furthermore, the proposed correlations can be used to predict bulk properties, such as the gas--liquid pressure drop across a PBR at microgravity conditions, but not interstitial properties, such as the flow regime, which are limited to bubble and pulse flow in NASA's PBRE at microgravity conditions. 

\begin{figure*}
    \centering
    \begin{subfigure}[t]{\linewidth}
      \centering
      \includegraphics[width=0.85\linewidth]{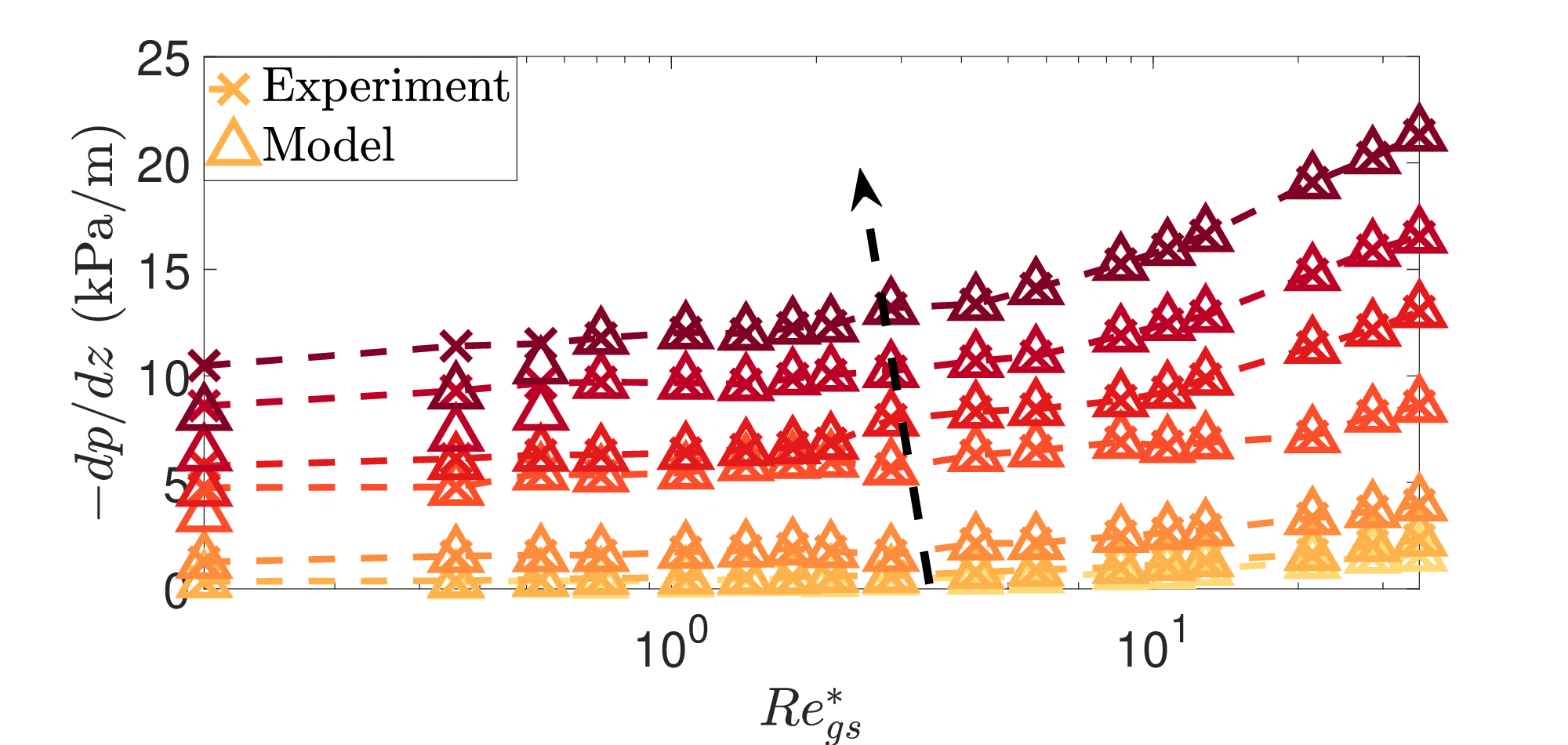} 
      \caption{}
      \label{fig:model_val_teflon}
    \end{subfigure}
    \begin{subfigure}[t]{\linewidth}
      \centering
      \includegraphics[width=0.85\linewidth]{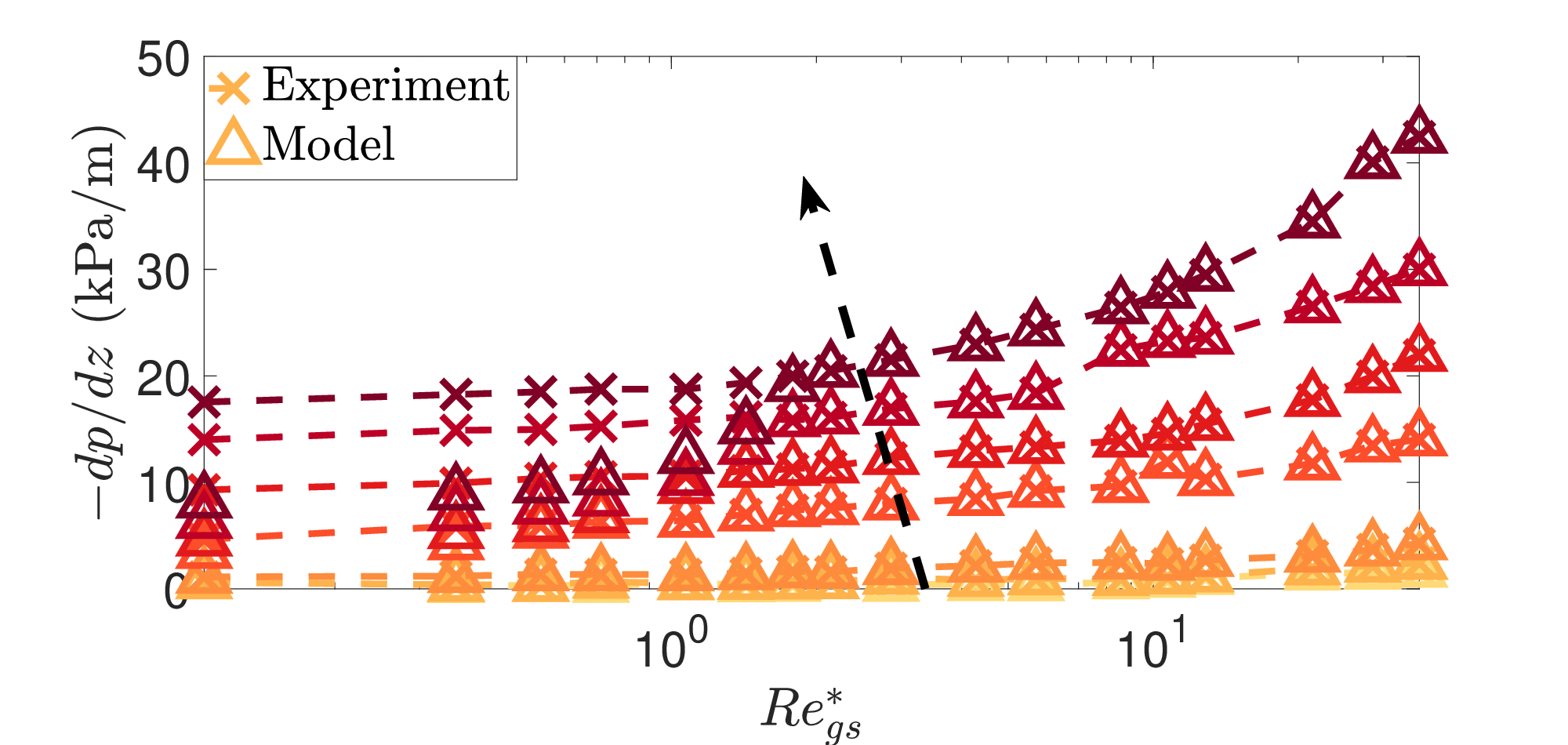}
      \caption{}
      \label{fig:model_val_glass}
    \end{subfigure}
    \caption{Verification of the proposed $f_{gl}$ correlations in Eq.~\eqref{eqn::f_gl} (and hence the approach) by comparing the model and experiment's $-dp/dz$ variation with $Re^*_{gs}$ at different $Re^*_{ls}$ for (a) Teflon packing and (b) glass packing PBRE data. As expected, many of the model predictions ($\bigtriangleup$) at discrete data points coincide exactly with the values from experiments ($\times$). The arrow indicates the direction of increasing $Re^*_{ls}$. The dashed lines are a guide to the eye and do not represent any trends. Note the logarithmic abscissa.}
    \label{fig::model_val}
\end{figure*}

\subsection{Model Verification}
\label{sec:model_val}
In this subsection, we verify (i.e., perform a \emph{self-consistency check} on) the proposed $f_{gl}$ correlations given in Eqs.~\eqref{eqn::f_gl}. We used $\epsilon=0.345$ and $d_p=3~\si{\milli\meter}$ in our calculations to be consistent with the PBRE dataset (i.e., the experiments of \citet{Motil2021GasliquidExperiment}). Specifically, we resolve Eq.~\eqref{eq::colm_combined}, which was obtained by eliminating the pressure gradient from the momentum equations~\eqref{eqn::colm_assumption}, by treating $\beta$ (in $f_{gl}$) as one of the inputs (along with $v_{ls}$ and $v_{gs}$) and using \textsc{Matlab}'s \texttt{fsolve} routine (with default settings, such as tolerances) to solve the nonlinear equation~\eqref{eq::colm_combined}, which yields $\phi_l$ as an output. The process is repeated for each $Re^*_{gs}$ and $Re^*_{ls}$. Then, $-dp/dz=A_{ls}/\phi_l^2$ (recall Sec.~\ref{sec:cons_law}), which is compared to the PBRE dataset values of $-dp/dz$ in Fig.~\ref{fig::model_val}. As seen from  Fig.~\ref{fig::model_val}, the pressure drops for both the packing materials match well, as expected, because the PBRE $-dp/dz$ data was used to infer $\beta$. Nevertheless, we observe a small number of outliers at large values of $Re_{ls}^*$ and low values of $Re_{gs}^*$. We hypothesize the mismatch for these few points is because the assumptions undertaken to develop the model, such as steady state and fully developed flow, may not hold in this regime with potentially unstable or unsteady flow existing, thus even though the value of $\beta$ (and hence $f_{gl}$) was fit to the data, the 1D TFM model does not reproduce it. In the present study, we utilize the proposed correlations in the limit of low $Re_{ls}^*$ and low $Re_{gs}^*$ where the model predictions are accurate. This comparison verifies the self-consistency of the approach by which we obtained the $f_{gl}$ correlations. Based on this approach, the next section provides a validation of the correlations.

\section{Example Implementation of the Proposed \texorpdfstring{$f_{gl}$}{f gl} in an Euler--Euler CFD Simulation}
\label{sec::cfd_model}
One approach to CFD simulation of a PBR is to explicitly resolve the gas--liquid flow through the tortuous paths created in a three-dimensional (3D) packing  \citep{Nagrani2024HydrodynamicsReactors,Ambekar2022ForcesBeds,Ambekar2021Pore-resolvedMap,Ambekar2022Particle-resolvedSize,Xu2022Particle-resolvedFlow}. Interface-resolved CFD simulations are computationally expensive, in particular because they require a fine mesh to resolve the interstitial flow domain between the packing structures. Another approach to CFD simulations of PBRs is an Euler--Euler two-fluid formulation \citep{Jakobsen2008ChemicalModeling}. In this approach, suitable drag force closures must be provided to a TFM to capture the liquid--solid and liquid-gas interactions. When properly calibrated, Euler--Euler simulations can be used to predict bulk (or integrated) quantities, such as the pressure drop across the PBR, but not the interstitial dynamics or flow regimes. However, the accuracy of this approach requires suitable drag correlations. In this section, we demonstrate that the correlations derived in Sec.~\ref{sec::math_model} from the PBRE data are useful for Euler--Euler CFD formulations.

\subsection{Simulation Methodology}
\label{sec:comp_method}
We simulated transient two-phase gas--liquid flow through an empty 2D planar channel using an Euler--Euler CFD formulation in ANSYS Fluent 2022 R1 \citep{ANSYS_Fluent1}. This formulation is based on a TFM. We account for porous medium (and hence porosity) implicitly by incorporating the $f_{gl}$ and $f_{ls}$ closure relations from Sec.~\ref{sec::math_model} using user-defined functions (UDFs).

\begin{figure*}[t]
    \centering
    \includegraphics[width=0.85\linewidth]{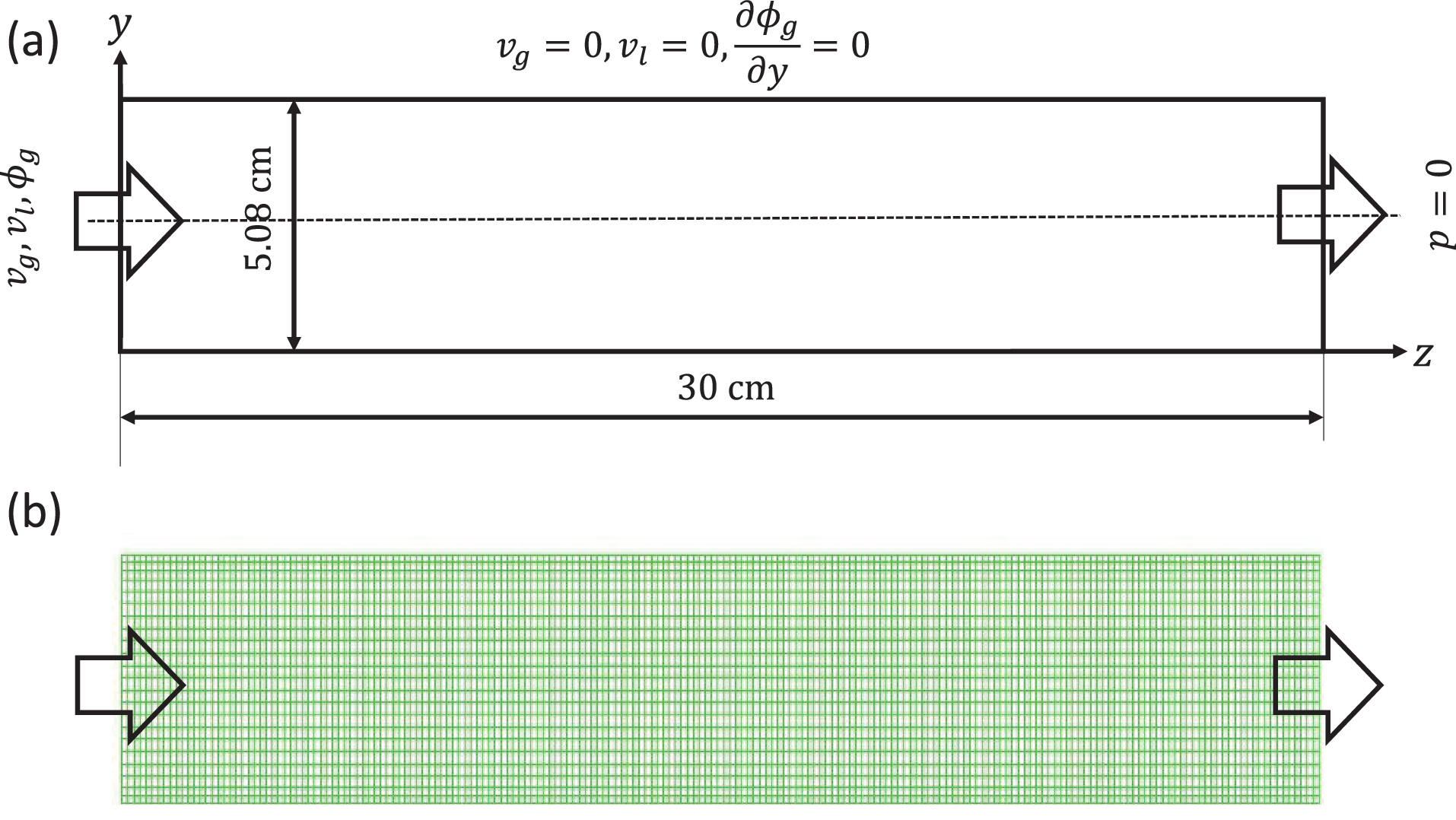}
    \caption{(a) Schematic of the 2D channel highlighting dimensions and boundary conditions and (b) mesh for the CFD simulations.}
    \label{fig::geom_mesh}
\end{figure*}

In the context of a TFM, which by definition does not resolve the details of the interstitial gas--liquid dynamics, the flow in a 3D PBR would be treated as axisymmetric, with suitable drag terms to take into account the liquid--solid and liquid-gas interactions. A 3D axisymmetric flow is essentially 2D, with an axial and cross-section coordinate. Thus, for simplicity but without loss of generality, we simulated a 2D channel geometry to test our approach to fitting the drag terms. The 2D channel and the corresponding mesh are shown in Fig.~\ref{fig::geom_mesh}. We created a CFD geometry similar to the experimental test facility for NASA's PBRE\citep{Motil2020PackedPBRE2,Motil2021GasliquidExperiment}. To reduce the computational power required to perform the simulations, however, the 2D channel width is $5.08~\si{\centi\meter}$ and its length is $30~\si{\centi\meter}$, which is half of the experimental setup's length \citep{Motil2021GasliquidExperiment}. We believe the flow is fully developed much upstream of this $30~\si{\centi\meter}$ region, and our CFD simulations predict linear pressure variation with axial position. Hence, the pressure drop from the CFD simulations can be compared to the data of \citet{Motil2021GasliquidExperiment}. In addition, the gas phase is nitrogen, the liquid phase is water, and constant properties are assumed as indicated in Table~\ref{tab::props}. 

\begin{table}
    \setlength{\tabcolsep}{5pt}
    \begin{tabular}{lccc}
      \hline
      \hline
      \noalign{\vskip 0.5ex}
      Phase & $\rho~(\si{\kilo\gram\per\meter\cubed})$ & $\mu~(\si{\pascal\second})$ & $\sigma~(\si{\newton\per\meter})$\\
      \hline
      \noalign{\vskip 0.5ex}
      Nitrogen & $1.165$ & $1.66\times10^{-5}$ 
      & \multirow{2}{*}{0.072} \\
      Water & $998.0$ & $1.03\times10^{-3}$ &  \\
      \hline
      \hline
    \end{tabular}
    \caption{Density $\rho$, dynamic viscosity $\mu$, and mutual surface tension $\sigma$ of nitrogen (gas phase) and water (liquid phase) used in the CFD simulations.}
    \label{tab::props}
\end{table} 

The boundary conditions (BCs) imposed on the CFD simulations are shown in Fig.~\ref{fig::geom_mesh}. At the inlet, constant mesoscale velocities calculated from the superficial velocities \citep{Gunjal2005HydrodynamicsModeling} are specified as $v_g(y,z=0,t)=v_{gs}/\phi_g(y,z=0,t)$ and $v_l(y,z=0,t)=v_{ls}/\phi_l(y,z=0,t)$ for all $t\ge0$ and $y$, assuming $\epsilon=1$ at the inlet plane. Since \citet{Motil2021GasliquidExperiment} do not report the inlet volume fraction for the phases, we assumed perfectly mixed conditions and assigned $\phi_g(y,z=0,t)=0.5$ for all $t\ge0$ and $y$. A zero pressure gradient is assigned at the outlet. No-slip velocity conditions ($v_g = v_l = 0$) and $\partial\phi/\partial y=0$ are imposed at the channel's top and bottom walls. As mentioned before, the porous medium is accounted for implicitly via the interphase coupling drag terms, $f_{gl}$ from Eq.~\eqref{eqn::f_gl} and $f_{ls}$ from Eq.~\eqref{eq::f_ls}. The source terms for each phase's axial momentum equation are defined as
\begin{subequations}\begin{align}
    f_{g} &= -f_{gl},\\
    f_{l} &= f_{gl}-f_{ls},
\end{align}\label{eqn::cfd_source_terms}\end{subequations}
and are implemented in ANSYS Fluent using a UDF based on local velocities (instead of the superficial velocities used in Eq.~\eqref{eq::f_ls} and Eq.~\eqref{eqn::f_gl}), following previous CFD studies \citep{Atta2007PredictionCFD,Atta2007InvestigationCFD,Lopes2008Three-dimensionalReactor,Lu2018ABeds,Gunjal2005HydrodynamicsModeling}.
We assumed constant porosity of $\epsilon=0.345$. For initial conditions, we assumed partial liquid holdup and hence $\phi_g(y,z,t=0)=0.5$ for all points $(y,z)$ in the domain. We also simulated different initial conditions, for example, $\phi_l(y,z,t=0)=0.8$, and found the final CFD pressure drop to be generally independent of the initial liquid holdup $\phi_l(y,z,t=0)$.  

In ANSYS Fluent \citep{ANSYS_Fluent1}, we used a ``Least Squared Cell Based" numerical scheme to discretize gradients. The pressure equation was discretized using the ``PRESTO!" scheme and ``First Order Upwind" schemes were used to discretize the momentum and volume fraction equations, while a ``First Order Implcit" transient formulation was employed to ensure the stability of the solver. Finally, the pressure-velocity coupling was achieved by the ``Phase Coupled SIMPLE" algorithm. To ensure the stability and convergence of the solver, we treated the source terms semi-implicitly, meaning the constant part of the source term is explicit, and the terms involving the velocity are implicit in the momentum equations (see Ref.~\citenum{ANSYS_Fluent2} for detailed explanation). 

A grid convergence study was conducted with three different grid sizes, with $1300$, $4378$, and $9600$ total number of elements. The final pressure drop across the 2D channel was found to be $\approx 4.3~\si{\kilo\pascal}$ in all three cases for the Teflon packing at the same inlet velocities. Thus, we employed the grid with $4378$ elements for the remaining simulations. 

\subsection{Comparison of Pressure Drop in PBRs}
\label{sec:pressure_drop}
We performed CFD simulations for Teflon and glass packings, each across different $Re^*_{gs}$ and $Re^*_{ls}$ values. Next, we compare the pressure drop from the CFD simulations to the experimental pressure drops in the PBRE dataset. We restrict our attention to the laminar regime of low $Re^*_{gs}$ and $Re^*_{ls}$ because we did not incorporate turbulence modeling in the simulations. However, it must be noted that we use the entire dataset \citep{Motil2020PackedPBRE2} to calculate $\beta$ and obtain the relevant $f_{gl}$ correlations (recall Sec.~\ref{sec::math_model}). The CFD simulations run until $\phi_l$ and $-dp/dz$ stop varying with time, and hence, a steady state has been established.

\begin{figure*}
    \begin{subfigure}[t]{\linewidth}
      \centering
      \includegraphics[width=0.85\linewidth]{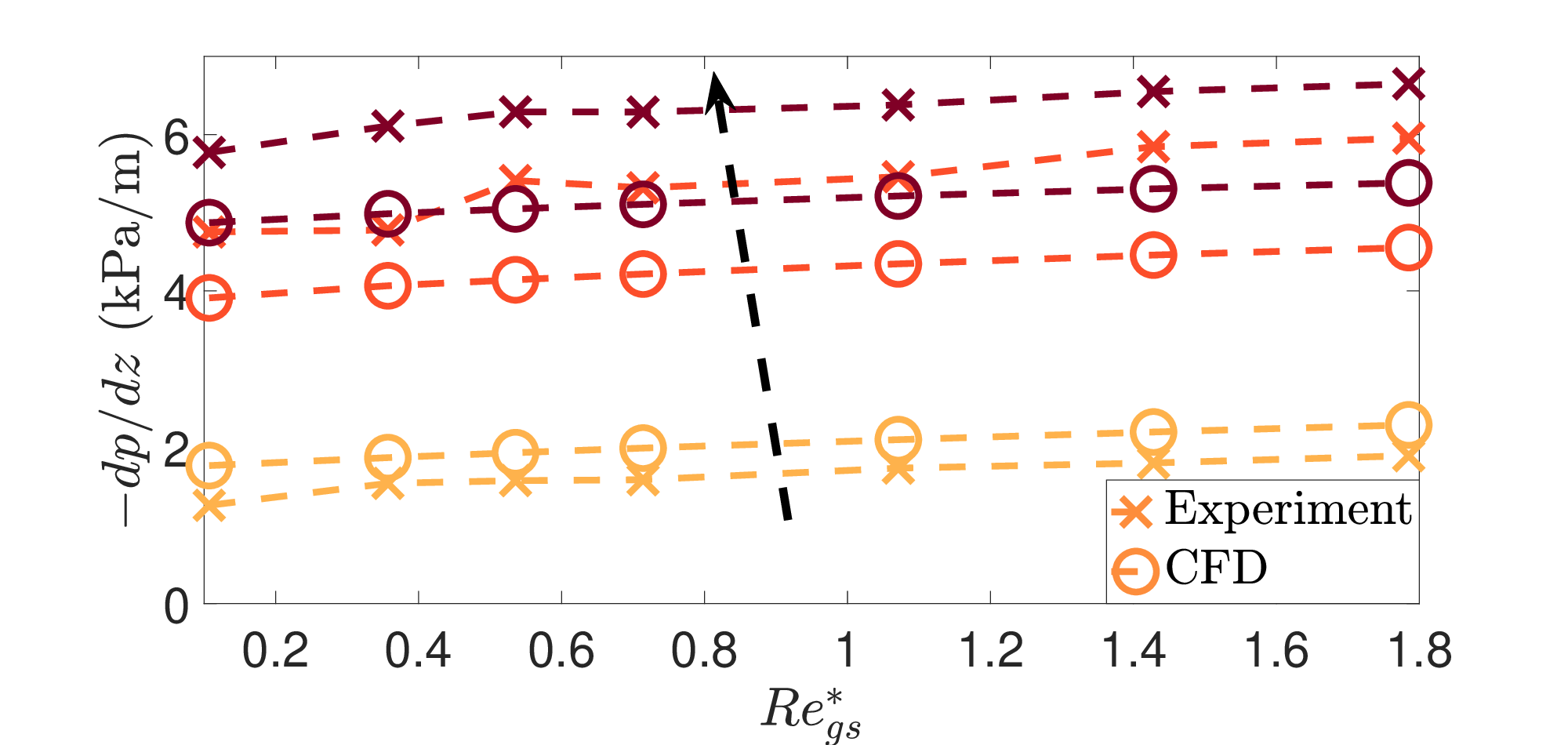}
      \caption{}
      \label{fig:model_cfd_teflon}
    \end{subfigure}
    \begin{subfigure}[t]{\linewidth}
      \centering    
      \includegraphics[width=0.85\linewidth]{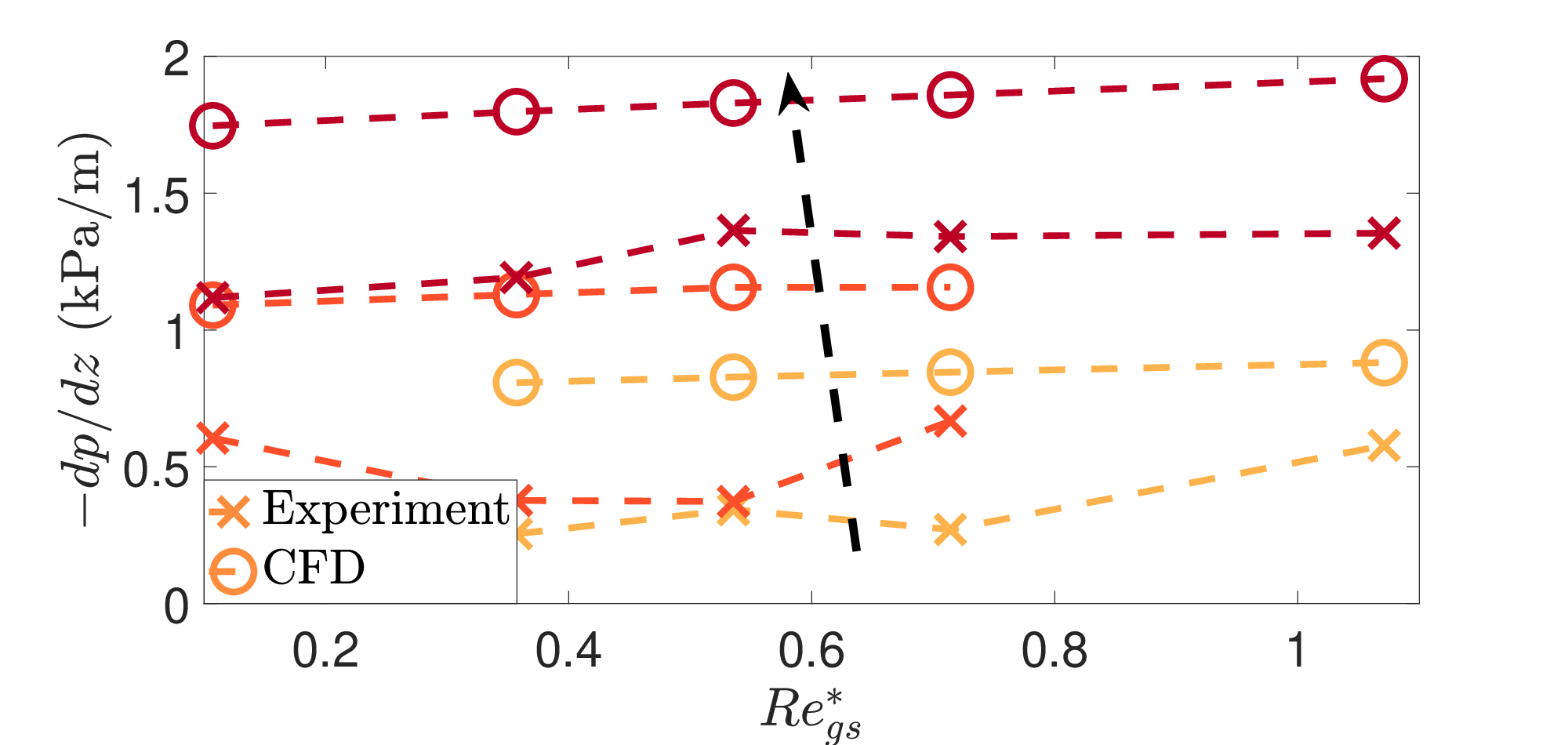}
      \caption{}
      \label{fig:model_cfd_glass}
    \end{subfigure}    
    \caption{Variation of CFD ($\bigcirc$) and experimental ($\times$) pressure gradients, $-dp/dz$, with the gas-phase Reynolds number, $Re^*_{gs}$, at three different liquid-phase Reynolds numbers $Re^*_{ls}$ (see the main text) for (a) Teflon packing and (b) glass packing. The arrow indicates the direction of increasing $Re^*_{ls}$. The dotted lines are a guide to the eye and do not represent any trends.}
    \label{fig::model_cfd}
\end{figure*}

Figure~\ref{fig::model_cfd} shows the variation of $-dp/dz$ with $Re^*_{gs}$ at different $Re^*_{ls}$ for (a) Teflon and (b) glass packing conditions. For the Teflon packing, in Fig.~\ref{fig:model_cfd_teflon}, the CFD simulation is performed for $Re^*_{ls} = 21.0$, $56.0$, $70.0$, and $0.1<Re^*_{gs}<2$. The CFD simulations predict a $-dp/dz$ within $\approx 25\%$ of the experimental values, highlighting the validity of the proposed approach based on the Euler--Euler simulation using the proposed $f_{gl}$ correlation for the Teflon packing. The highest error of $\approx 40\%$ occurs at the lowest flow rates and is an artifact of the small pressure drops under consideration. The PBRE datasets\citep{Motil2020PackedPBRE2} do not report the uncertainty of all experimental measurements. 

Similarly, for the glass packing, simulations are performed for $Re^*_{ls} =3.5$, $8.4$, $21.0$, and $0.1<Re^*_{gs}<1.1$ to restrict the simulations to the laminar flow regime. As seen in Fig.~\ref{fig::model_cfd}, the CFD simulations predict $-dp/dz$ values similar to the experimental ones. However, higher percent errors are observed due to the low $-dp/dz$ values for glass beads and also because of higher uncertainty in the experimental values, especially for small $Re^*_{ls}$. The uncertainty in the experiments for glass bead packing is expected to be large as there is a nonmonotonic behavior of $-dp/dz$ near $Re^*_{ls}=8.4$. It must be noted that the error, $\approx 25\%$ between CFD simulations and experimental measurements, is typical for such multiphase flows simulations of flow through PBRs \cite{Dixon2021Particle-resolvedNumber,Bai2009AParticles,Jurtz2020ValidationDynamics}. Indeed, previous studies report that even with particle-scale resolve simulations and idealized conditions, it is challenging to reduce the pressure drop error below 10\%.

Finally, \emph{bulk} quantities, such as the pressure drop across the PBR at steady-state, are expected to be fairly insensitive to wettability effects. This claim is supported by our simulations at $Re^*_{ls}=21.0$, for which the pressure gradient is approximately $1.75~\si{\kilo\pascal\per\meter}$ for \emph{both} glass and Teflon packings, therefore not significantly influenced by the different wettability of the packings. However, it must be noted that the \emph{local} (pore-scale) flow dynamics and behavior are strongly sensitive to the wettability \cite{Zhao2016WettabilityMicrofluidics}. 

\section{Conclusion}
\label{sec::Conclusion}
Motivated by the lack of studies of two-phase gas--liquid flow through a PBR at microgravity conditions, we used PBRE data from NASA's PSI system \citep{Motil2020PackedPBRE2}, which accesses only the bubble and pulse flow regimes at microgravity for monodisperse Teflon (nonwetting) and glass (wetting) packing materials, to develop a closure relation for $f_{gl}$ for two different packing materials. Based on the predominant approach in the literature, we assumed fully-developed 1D flow at steady-state with cross-sectionally uniform velocities and shared pressure gradients between the gas and liquid phases, thus rendering the TFM into a single Eq.~\eqref{eq::colm_combined}. Using the latter, we followed a data-driven methodology to calculate a drag coefficient $\beta$, featured in the gas--liquid drag, $f_{gl}=\beta (v_g-v_l)$, for each value of the gas and liquid Reynolds numbers, $Re^*_{gs}$  and $Re^*_{ls}$, available in the PBRE dataset. Then, we performed composite curve fitting to form the correlations for $f_{gl}$ for Teflon and glass packing, given in Eqs.~\eqref{eqn::f_gl}. The proposed interphase drag correlations are expected to apply for $0.1 \le Re^*_{gs}\le 35.7$ and $3.5 \le Re^*_{ls} \le 104.4$, the ranges of Reynolds numbers in the NASA PBRE datasets, which were used to optimize the drag in our model.

Next, we demonstrated the utility of the proposed $f_{gl}$ correlation within a CFD simulation. Specifically, we used an Euler--Euler formulation in ANSYS Fluent to predict the pressure drop across a PBR under microgravity conditions. The proposed closure relations were incorporated as sources in the axial momentum equations via UDFs. We performed CFD simulations in the laminar regime. The resulting pressure drops were compared to the experimental ones reported in the PBRE dataset. We observed general agreement between the two across different packing materials, within a tolerance of $\approx 25\%$.


\begin{acknowledgement}
This research was supported by the National Aeronautics and Space Administration under Grant No.\ 80NSSC22K0290.
Simulations were performed using the community clusters of the Rosen Center for Advanced Computing at Purdue University.
\end{acknowledgement}

\section*{Associated Content}
\subsection*{Data Availability Statement}
The data underlying this study are openly available in the Purdue University Research Repository (PURR) at \url{http://doi.org/10.4231/S34E-8N27}. 

Therein, the zip archive named ``Math\_model\_figs.zip'' contains the MATLAB scripts for implementing the model described in Sec.~\ref{sec::math_model}, and the code to make Figs.~\ref{fig::fgl_fit_teflon_all}, \ref{fig::fgl_fit_glass_all}, and \ref{fig::model_val}.
This zip archive also contains the pressure drop data downloaded from NASA's PSI system and the post-processed CFD data from the present study. The code to generate Fig.~\ref{fig::model_cfd} is also included. 
The zip archives named ``Glass\_packing.zip" and ``Teflon\_packing.zip" contain the CFD simulation case files along with the UDF for ANSYS Fluent, as described in Sec.~\ref{sec::cfd_model}.


\section*{Author Contributions}
\textbf{Pranay P.\ Nagrani}: conceptualization (equal); data curation (lead); formal analysis (lead); investigation (equal); methodology (equal); validation (lead); visualization (lead); writing - original draft (lead).
\textbf{Amy M.\ Marconnet}: writing - review \& editing (supporting); supervision (equal); funding acquisition (supporting).
\textbf{Ivan C.\ Christov}: conceptualization (equal); formal analysis (supporting); investigation (equal); methodology (equal); writing - original draft (supporting); writing - review \& editing (lead); supervision (equal); funding acquisition (lead).


\bibliography{pbr_1d_tfm_abbv}


\end{document}